\documentclass[aps,preprint]{revtex4}
\usepackage{amssymb}

\input{tcilatex}

\begin{document}

\preprint{}
\title{The $\mathcal{R}_{0}$ Approach to Epidemic-non-Epidemic Phases
Revisited}
\author{O.E. Aiello and M.A.A. da Silva}
\affiliation{Departamento de F\'{\i}sica e Qu\'{\i}mica da FCFRP, Universidade de S\~{a}o
Paulo, 14040-903 Ribeir\~{a}o Preto, SP, Brazil}

\begin{abstract}
In this work, we revisit the basic reproduction rate $\mathcal{R}_{0}$
definition for analysis of epidemic-non-epidemic phases describing the
dynamics of the discrete stochastic version of the epidemic $SIR$ model
based on the Master Equation formalism. One shows that it is a very precise
and efficient way to determine the epidemic threshold; using its most
primitive concept, we can find exact results.
\end{abstract}

\volumeyear{}
\volumenumber{}
\issuenumber{}
\eid{}
\date{February 13, 2003}
\received[Received text]{}
\revised[Revised text]{}
\accepted[Accepted text]{}
\published[Published text]{}
\startpage{01}
\endpage{}
\maketitle

\textit{Introduction} - The basic reproduction rate, $\mathcal{R}_{0}$, is
the most fundamental parameter used by epidemiologists \cite{Gani}. It has
raised interest of physiscists because of analogies between epidemic and
percolation systems \cite{Warren,Dickman,Newman,Benedita}, and its
generality, that permits, for example, analyze spreading of viruses in a
structured scale-free network \cite{Victor}. The definition of $\mathcal{R}%
_{0}$ is ``the average number of secondary cases caused by an infectious
individual in a completely susceptible population'' \cite{Anderson92}. This
simple idea had a profound effect on epidemic theory. It may be a global
insight that cuts through the details of the transmission process, because
it originated from consideration of deterministic models of homogeneous
population with random mixing \cite{Bailey}. However, putting more details
in the model one can extend its definition to heterogeneous mixing. In this
way, naturally the efforts were in analytical calculations of $\mathcal{R}%
_{0}$ for continuous deterministic or stochastic models \cite%
{Warren,Jacquez,Keeling}. The stochastic framework, although more realistic
in principle, it is more complex to analyze because of detail required \cite%
{Mollison,Murray,Alves}. Commonly, simulations helped to confirm theoretical
assumptions. Improved machine technology has spread the use of
computationally intensive methods to solve a great diversity of epidemics,
and so simulation technics, as Dynamical Monte Carlo (DMC) \cite%
{Binder2,Fichtorn,Gillespie,Aielo}, are becoming more popular in this
subject. Taken on the advantage of the DMC method, we can calculate $%
\mathcal{R}_{0}$ straightforward. Thus, we adopted this approach to
characterize the separation between the epidemic and non-epidemic phases,
because its efficiency and simplicity that enables a facile finding of exact
results.

Throughout this Letter, we shall consider the classical $SIR$ (Susceptible,
Infected, Removed) epidemic model, originally based on the chemical \ \
``mass action'' principle (see \cite{Haas} and references therein), to
illustrate the discrete stochastic approaches. Our model encloses the
deterministic one\ as a \ particular case. Based on the Master equation
formalism we generalize the $SIR$ model and $\mathcal{R}_{0}$ for discrete
stochastic systems. Also, we describe the local epidemic model with an exact
result to $\mathcal{R}_{0}$ using its primitive concept. Finally, we show a
phase diagram of the $local-SIR$ model, with deterministic-like behavior,
where no homogeneous mixing is considered, and dynamical Monte Carlo
simulation results.

\textit{Generalized SIR model and }$R_{0}$ - Stochastic process approaches
could simulate non-equilibrium systems, even the deterministic ones,
introducing random variables to describe them in a microscopic scale. The
macroscopic behavior of some system is resulting from averages of its
microscopic properties. One can describes the evolution of the distribution
of probabilities, for markovian processes, with the \emph{Master Equation}: 
\begin{equation}
\frac{dP_{i}(t)}{dt}=\sum_{j}w_{j\rightarrow
i}P_{j}-\sum\limits_{j}w_{i\rightarrow j}P_{i},
\label{Pauli Master Equation}
\end{equation}%
where $P_{i}$ is the probability to find the system at the state $i$ at the
time $t$, and $w_{i\rightarrow j}$ is the transition probability per unity
of time. Considering $T_{ij}$ the probability of transition from $i$ to $j$,
we may write $w_{i\rightarrow j}=\frac{T_{ij}}{\tau _{i}}$ \ \cite{Hoel},
where $\tau _{i}$ is a characteristic time constant (\textit{lifetime}) of
the state $i$.

We now start by choosing a convenient physical extensive microscopic
quantity $A_{i}$, which depends only of the system's state $i$. Since the
time must change for every successful event, we will consider only counting
events related quantities. To $SIR$ epidemic systems the number of infected
individuals, for example, is an adequate quantity because it represents the
balance between the number of infection and removal events. The mean value
for a given quantity at the time $t$ is 
\begin{equation}
A(t)=\langle A\rangle =\sum_{i}P_{i}(t)A_{i}.  \label{Temporal Mean Value}
\end{equation}%
This equation represents a continuous physical macroscopic quantity $A(t)$.
We can differentiate both sides of the equation above, with respect to $t$.
After that, using $\left( \ref{Pauli Master Equation}\right) $, and by
defining $\Delta A_{ij}=A_{i}-A_{j}$, we get 
\begin{equation}
\frac{dA(t)}{dt}=\sum_{i}\sum_{j}w_{j\rightarrow i}P_{j}\Delta A_{ij}.
\label{Macroscopic Master Equation 2}
\end{equation}

Consider now the nearest-neighbor states $j$ of a given state $i$; if we
measure the \ ``distance'' between the states, say by the quantity $|\Delta
A_{ij}|$, such that the non-null minimum value is $|\Delta A_{ij}|=a$, we
may approach the equation $\left( \ref{Macroscopic Master Equation 2}\right) 
$ by:%
\begin{equation}
\frac{dA(t)}{dt}=\sum_{(ij)}w_{j\rightarrow i}P_{j}a\delta _{ij},
\label{Macroscopic Master Equation 3}
\end{equation}%
where the symbol \ $(ij)$ denotes a nearest-neighbor pair of states, and $%
\delta _{ij}=\Delta A_{ij}/|\Delta A_{ij}|$. Now we consider another
physical quantity $A^{\dagger }$ that represents a source for the quantity $%
A $. Thus, we can rewrite $\left( \ref{Macroscopic Master Equation 3}\right) 
$ as:%
\begin{equation}
\frac{dA(t)}{dt}=\sum_{j}r_{j}^{+}P_{j}A_{j}^{\dagger
}-\sum_{j}r_{j}^{-}P_{j}A_{j},  \label{Macroscopic Master Equation 4}
\end{equation}%
where $r_{j}=<w_{j\rightarrow i}>_{i}$ are the averaged transition
probabilities per unity of time over the ensemble of the nearest-neighbor
states $i$ of $j$ at some time $t$, i.e., the \textit{mesoscopic} rates.
Here, the word ensemble means a set of configurations accessible at a finite
(small) time around a time $t$; in this sense we are using a time dependent
ergodicity idea \cite{Binder2}, and so generally the systems are \ non
ergodic in nonequilibrium states. The \ superscripts $\ ``+"$ and $\;``-"$
mean respectively the contributions to increasing and to decreasing the
quantity $A(t)$ \cite{Aielo2}.

Based on $\left( \ref{Macroscopic Master Equation 4}\right) $, we formulated
the $GSIR$ model through the following set of stochastic differential
equations and inter-classes rates:%
\begin{eqnarray}
\frac{dS}{dt} &=&\sum_{j}r_{RS}^{j}P_{j}R_{j}-\sum_{j}r_{SI}^{j}P_{j}S_{j},
\label{DSDT} \\
\frac{dI}{dt} &=&\sum_{j}r_{SI}^{j}P_{j}S_{j}-\sum_{j}r_{IR}^{j}P_{j}I_{j},
\label{DIDT} \\
\frac{dR}{dt} &=&\sum_{j}r_{IR}^{j}P_{j}I_{j}-\sum_{j}r_{RS}^{j}P_{j}R_{j}.
\label{DRDT}
\end{eqnarray}%
The mesoscopic rates are $r_{SI}^{j}$ , $r_{IR}^{j}$ and $r_{RS}^{j}$, for
each state $j$, respectively, from $S\rightarrow I$, $I\rightarrow R$ and $%
R\rightarrow S$. To satisfy the $SIR$ condition the set $\{r_{RS}^{j}\}$ is
null. Note that we meant that, for example, if $A=I$, then $A^{\dagger }=S$
in the equation $\left( \ref{Macroscopic Master Equation 4}\right) $. The
conservation law with the total number of individuals $N=S(t)+I(t)+R(t)$ is
satisfied. One may obtain the reproduction rate, $\mathcal{R}_{0}$, directly
from the equations $\left( \ref{DSDT}-\ref{DRDT}\right) $ with the epidemic
condition $\frac{dI}{dt}\geq 0;$ where the equality is the threshold and it
is set to $t_{0}=0,$ the initial time. Thus, we can do $%
\sum_{j}r_{SI}^{j}P_{j}S_{j}-\sum_{j}r_{IR}^{j}P_{j}I_{j}\geqslant 0$, what
implies that $\sum_{j}r_{SI}^{j}P_{j}S_{j}\geqslant
\sum_{j}r_{IR}^{j}P_{j}I_{j}$. One can thus write the reproduction rate as%
\begin{equation}
\mathcal{R}_{0}=\frac{\sum_{j}r_{SI}^{j}P_{j}S_{j}}{%
\sum_{j}r_{IR}^{j}P_{j}I_{j}},  \label{R0-geral}
\end{equation}%
for stochastic processes. As the condition to $\mathcal{R}_{0}$ to the
epidemic threshold must be valid to the ensemble average of initial states $%
j_{0}$ that gives the same initial condition $S_{0},$ $I_{0}$ and $R_{0}$,
we may define%
\begin{equation}
\mathcal{R}_{0}=\frac{<r_{SI}>_{0}S_{0}}{<r_{IR}>_{0}I_{0}};  \label{R0}
\end{equation}%
the average number of the secondary cases produced by $I_{0}$ infected
initially. Note that if we do $<r_{SI}>_{0}=bI_{0}$ and $<r_{IR}>_{0}=a$ \
we recover the\ deterministic case \cite{Murray}. One must observe that if
some initial configuration is fixed, one does not need the averages in $%
\left( \ref{R0}\right) $, but only obtain the rates, $r_{SI}$ and $r_{IR}$;
so, of course, generally, the $\mathcal{R}_{0}$ depends on the initial
configuration choice. In many practical situations this is important because
it will determine an epidemic or not. We can easily adapt the result above
to other models, as the $SIS$ model \cite{Victor}, for example, to analysis
of the epidemic threshold.

\textit{Local epidemic model} - Generically, the temporal and spatial
evolution characterize any epidemics, where in each part of the system the
density of the elements can vary with the time. One can analyze this process
through a two-dimensional lattice, in which each site, representing an
individual of the population, receives own attributes as susceptibility and
interactivity referring each site with the others. We will analyze a model
with local contact only. The elements are all fixed, i.e., no populational
mobility is considered. The main reason to study a such particular model is
that we have more fluctuations, and so it is a good test to the efficiency
of our approach.

The probability of individuals contracts the illness, in transmitting a
disease by contact, depends on the \emph{status} (susceptible, infected or
removed) in which they meet its neighbors; its chance of getting sick will
depend on the number of sick neighbors. Thus, considering an element
possesses $n$ infective neighbors, and an infection chance, $p_{0}$, due to
each neighbor, the probability of its change in a sick element (through $n$
effective contacts) will be \cite{Haas}:%
\begin{equation}
w_{_{SI}}=\Lambda \lbrack 1-(1-p_{0})^{n}]\text{.}  \label{p}
\end{equation}%
Therefore, $(1-p_{0})^{n}$ is the probability of no infection of a
susceptible (individual) if it has $n$ infected neighbors, thus $%
1-(1-p_{0})^{n}$ is the probability of infection of a susceptible if it has $%
n$ infected neighbors. The $\Lambda $ parameter gives the $w_{_{SI}}$ as
inverse of time units. When $n=0$, that is, when no neighbor is
contaminated, the probability of contamination due to contact is zero, so $%
w_{_{SI}}$ increases when the number of effective contacts, $n$, increases.
A global removal rate determines the infectious period, and an infected
individual turn immune (removed) stochastically. So the infectious period
for each individual fluctuates over an average number given by the inverse
of the removal rate, like in the mean field approach \cite{Anderson79}. In
this sense our definition of $\mathcal{R}_{0}$ is more general than that
gave in the reference \cite{Alves}. Also has a difference that it is
considered instantaneously instead during the infectious period, so it
follows close the classical definition. However, we considered a range of
initial values to the number of infected individuals, and did an analysis
with an initial random distribution of immunes. Note that for the considered
model in a square lattice, when $I_{0}=1$, we have the trivial exact result%
\begin{equation}
\mathcal{R}_{0}=\frac{n_{\max }\Lambda p_{0}}{w_{IR}},  \label{R0Local}
\end{equation}%
where $n_{\max }$ is the maximum number of contacts. One can see that the
exact result to a more thorough model, including the homogeneous mixing
(mean field), is straightforward. The reference \cite{Alves} shows this
result as an analytical approximation to $\mathcal{R}_{0}$.

\textit{Results and final remarks} - For practical purposes one distributes
the individuals on a square lattice of $N=M\times M$ sites. All the
individuals at the lattice boundary have their \emph{statuses} fixed at
susceptible \emph{status}. One considered only two lattice sizes, $M=10$ and 
$200$, because increasing $M$ reaches smooth curves near to the $M=200$
rapidly. We did $r_{IR}^{j}=q=1$, constant independent of the configuration,
and $r_{SI}^{j}=\ <w_{SI}>_{j}$, averaged over all individual probabilities
that appear in any random configuration $\ j;$ with $w_{SI}$ modeled with a
purely local interaction with $\Lambda =1$, so $w_{SI}$ $=1-(1-p_{0})^{n}$.
The initial condition for the number of infectives, $I_{0}$, to the system,
for $M=10$, was varying from $1$ to $99$, and, for $M=200$, from $1$ up to $%
38000$ randomly distributed on the lattice. One occupies the remaining sites
by $S_{0}=N-I_{0}$ susceptibles, so $R_{0}=0$ for both cases. We did the
variable $n$ \ as an integer ranging from $0$ to $8$, since the first and
second nearest infected neighbors are indistinguishably considered for each
susceptible.

Considering that the sum of the ``microscopic influences'' creates a rate $%
r_{SI}$, we calculated by a sample average the initial mean rate $%
<r_{SI}>_{0}$. Of course in the practice of the simulation we drew randomly
only a few configurations to estimate the averages and we used the equation $%
\left( \ref{R0}\right) $ to elaborate the phase diagrams for the $SIR$ model
showed in the figure 1. We got the values for $<r_{SI}>_{0}$ using $100$ and 
$4\times 10^{4}$ configurations for $M=200$ and $10$ respectively. Observe
that the maximum $\mathcal{R}_{0}$ happens when the probability $p_{0}$
reaches its largest value in a great population of susceptible. We found for 
$M=10$, the epidemic threshold in the interval $\mathcal{R}_{0}=1.03-1.05$.
For $M=200$, the interval reduced to a very thin line defined at $\mathcal{R}%
_{0}=1$, as expected to infinite systems. Note that the $\mathcal{R}_{0}$
contour lines for $M=10$ are noisy because of the small size of the lattice.%
\emph{\ }For cases when $\mathcal{R}_{0}>1$, happen epidemic bursts \emph{in
average}, and when $\mathcal{R}_{0}<1$ does not. We did, also, experiments
with a non-zero initial immune individuals number, $R_{0}$, and its
qualitative effect on $\mathcal{R}_{0}$ is the same as for $R_{0}=0$, since
we have a random initial distribution of immunes. Quantitatively, for the
initial infectives number fixed, we need to increase the probability per
unit of time, $w_{SI}$, to have epidemics; as expected, has a critical value
to $R_{0}$ that no epidemic occurs, the so known herd immunity effect \cite%
{Aielo}. For the considered parameters of our model, we are at the epidemic
threshold when%
\begin{equation}
<r_{SI}>_{0}^{t}=\frac{I_{0}}{S_{0}}.  \label{mean initial SI rate}
\end{equation}%
After several fittings we found the reasonable expression to the initial
average rate: $<r_{SI}>_{0}^{t}=\alpha \lbrack 1-(1-p_{0})^{\beta }]$, with
the real numbers $\alpha \leq 1$ and $0\leq \beta \leq 8$. The critical
value for $R_{0}$ is so: $R_{0}=N-I_{0}[1+1/<r_{SI}>_{0}^{t}]$. Note that if
the condition $I_{0}>\alpha S_{0}$ is satisfied we have no epidemics for any
value of $p_{0}$. At first sight it might be a strange result, however, if
the initial state has most infected individuals, the removal chance of some
is high; so it decreases faster than the number of infections itself.

The figure $2$ show some epidemic and non epidemic cases obtained by
Dynamical Monte Carlo. This direct method was used to find the threshold
line of the phase diagrams showed in the figure 1. The very beginning of the
process is sufficient to determine $\mathcal{R}_{0}$, but for completeness
we showed the epidemic curves in an extended time.

In summary, we presented a stochastic version for the parameter $\mathcal{R}%
_{0}$, based on the description of the $SIR$ model, by means of the Master
equation formalism. This way, the predictive power $\mathcal{R}_{0}$ is
transported from deterministic to the stochastic one, generalizing the
concept. In fact, a very defined result to the threshold curve was earlier
found to deterministic systems; it is interesting that we can have this,
also, for stochastic systems. It is consequence of the definition of $%
\mathcal{R}_{0}$, whose difference with that one of already cited recent
work \cite{Alves} did decrease the threshold. Fluctuations to favor
epidemics in our case when $\mathcal{R}_{0}<1$ are smaller than those favor
non-epidemics, i.e., the number of non-epidemic cases prevails, even for
finite small systems; the same happens with the opposite case. Complex
geometries can be included in this model, since have no restrictions to the
model in this sense, and the system geometry is an important factor to
change the threshold. If the system is strongly dependent on the initial
conditions, the averages are not appropriated to predict an epidemic. We
believe that this definition of $\mathcal{R}_{0}$ open the doors for new
investigations and calculations of $\mathcal{R}_{0}$ for more realistic
systems because we used a general microscopic description to get a parameter
of macroscopic nature.

The authors gratefully acknowledge funding support from FAPESP Grant n.
00/11635-7 and 97/03575-0. The authors would also like to thank Drs. A.
Caliri and V.J. Haas for many stimulating discussions and suggestions.

\newpage

FIGURE CAPTIONS

\bigskip

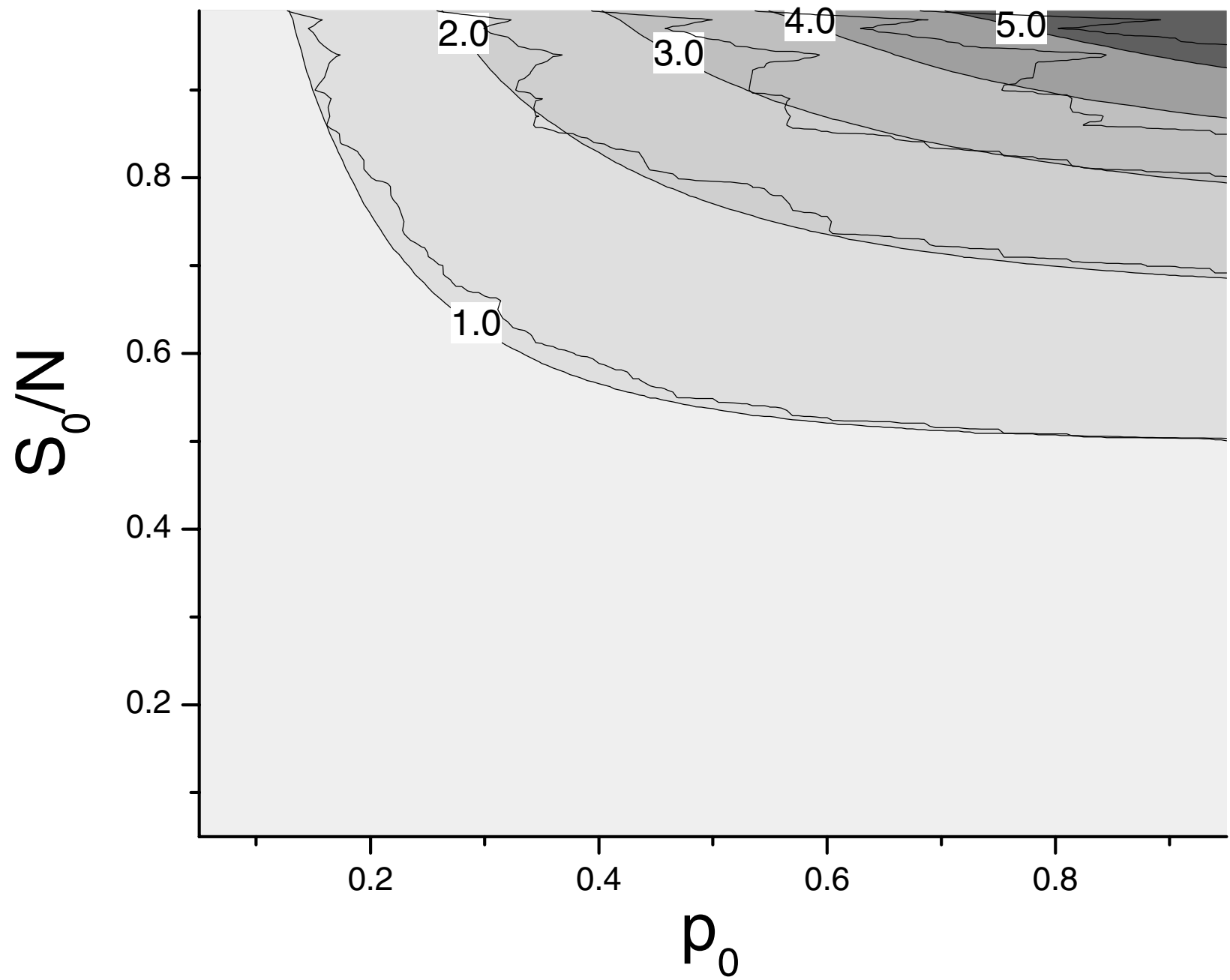
Figure 1. Shows a phase diagram for the $local-SIR$ model with parameters $%
q=1.0$, $\frac{S_{0}}{N}=(0.05-0.95)$ and $p_{0}=(0.05-0.95)$. The values of 
$\mathcal{R}_{0}$ larger than $1$ in the smooth contour lines ($M=200$)
allow that the number of infected increases with the time characterizing an
epidemic outbreak, for $\mathcal{R}_{0}$ smaller than $1$ the infection
fade-out. The threshold to the noisy contour map ($M=10$) is in the interval 
$1.03-1.05$ that is too small to show in the diagram.

\bigskip

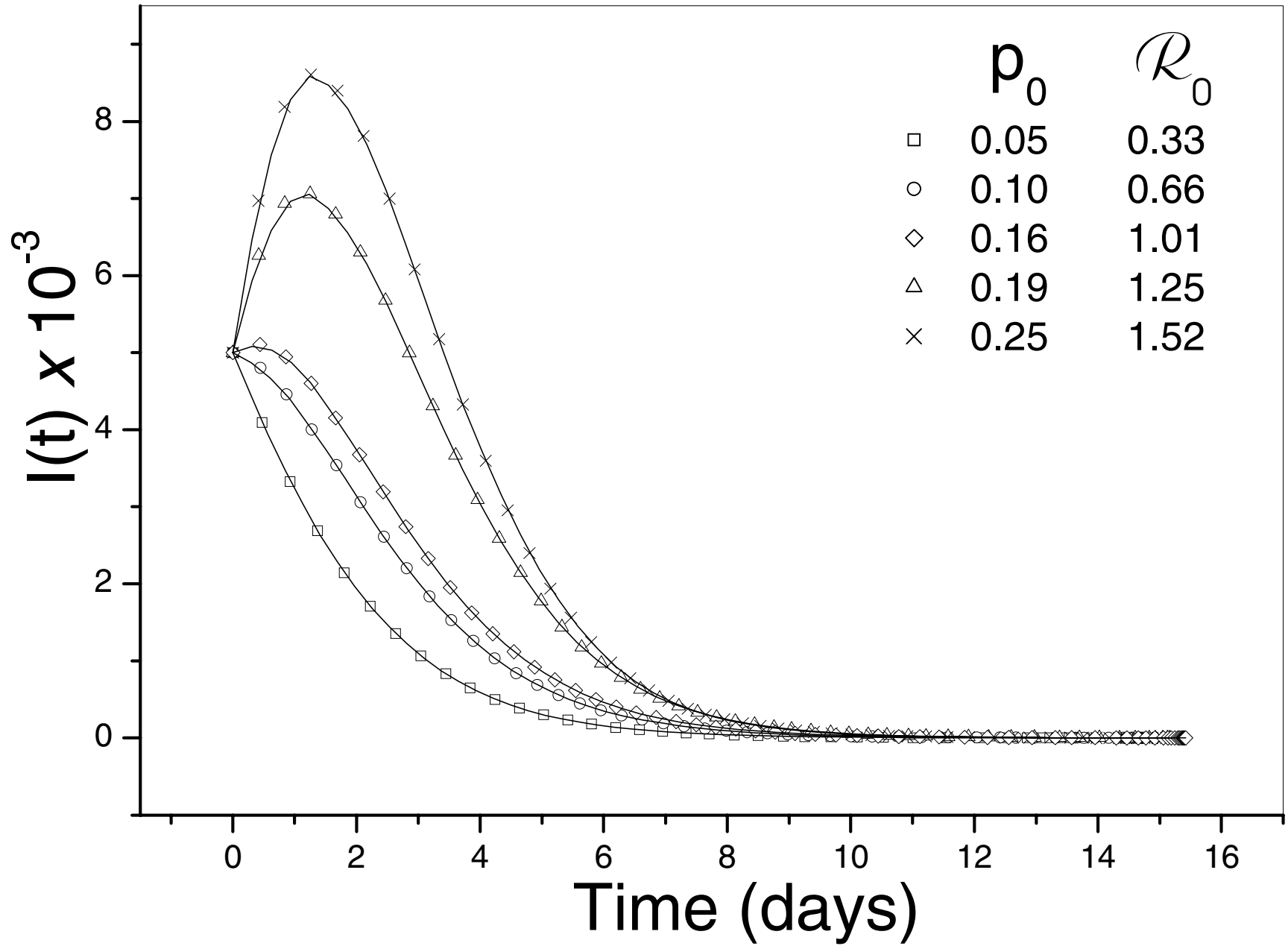
Figure 2. Epidemic curves. Show the number of Infectives evolving with the
time. The numerical values for the model parameters are $q=1$, $M=200$ and $%
I(0)=5000.$ A total of $20$ experiments was done to get the averages. The
figure shows two cases, some curves those represent epidemic outbreaks ($%
\mathcal{R}_{0}>1$) and others in that the infection fade-outs ($\mathcal{R}%
_{0}<1$).

\bigskip\

\end{document}